\documentclass[]{interact}

\usepackage{epstopdf}
\usepackage[caption=false]{subfig}

\usepackage{natbib}
\usepackage{color}
\usepackage{hyperref}
\bibpunct[, ]{(}{)}{;}{a}{}{,}
\renewcommand\bibfont{\fontsize{10}{12}\selectfont}

\theoremstyle{plain}

\theoremstyle{definition}

\theoremstyle{remark}

\begin{document}

\articletype{REVIEW}

\title{Computational models advance deep brain stimulation for Parkinson's disease}

\author{
\name{Yongtong Wu\textsuperscript{1,$\dagger$} , Kejia Hu\textsuperscript{2,3,$^\ast$,$\dagger$}\thanks{$^\ast$ CONTACT Kejia Hu. Email: dockejiahu@gmail.com; Shenquan Liu. Email:mashqliu@scut.edu.cn} and Shenquan Liu\textsuperscript{1,$^\ast$}\thanks{$\dagger$ These authors contributed equally to this work.}}
\affil{\textsuperscript{1}School of Mathematics, South China University of Technology, Guangzhou, Guangdong, China; \textsuperscript{2}Department of Neurosurgery, Ruijin Hospital, Shanghai Jiao Tong University School of Medicine, Shanghai, China;
\textsuperscript{3}Center for Functional Neurosurgery, Ruijin Hospital, Shanghai Jiao Tong University School of Medicine, Shanghai, China;
}
}

\maketitle

\begin{abstract}
Deep brain stimulation (DBS) has become an effective intervention for advanced Parkinson's disease, but the exact mechanism of DBS is still unclear. In this review, we discuss the history of DBS, the anatomy and internal architecture of the basal ganglia (BG), the abnormal pathological changes of the BG in Parkinson's disease, and how computational models can help understand and advance DBS. We also describe two types of models: mathematical theoretical models and clinical predictive models. Mathematical theoretical models simulate neurons or neural networks of BG to shed light on the mechanistic principle underlying DBS, while clinical predictive models focus more on patients' outcomes, helping to adapt treatment plans for each patient and advance novel electrode designs. Finally, we provide insights and an outlook on future technologies.
\end{abstract}


\maketitle

\section{Introduction}\label{sec1}

\subsection{Deep Brain Stimulation(DBS)}

Deep brain stimulation (DBS) is a developing technology used to treat various neurological diseases. It improves the patient's mobility and self-care by stimulating the relevant nuclei in the brain that control movement and "modulates" \citep{Ashkan2017} the abnormal brain neural signals through weak electrical pulses delivered by electrodes implanted in the brain. In 1987 Benabid and colleagues demonstrated that DBS produced beneficial effects similar to pallidotomy in patients with Parkinson's disease (PD) \citep{Benabid1987}. Moreover, compared to L-Dopa therapy \citep{Miyasaki2002} and pallidotomy \citep{Goetz2005}, DBS has considered one of the excellent options for PD relief today due to it is relatively minimally invasive, non-destructive, and reversible(compared to excisional surgery) \citep{Pahwa2006a}. Since then, the technique has gradually opened up new frontiers in the surgical treatment of various movement disorders, depression, obsessive-compulsive disorder, and Alzheimer's disease. By 2022, nearly 40 years after the development of DBS, the routine use of DBS for the treatment of the motor symptoms of PD has proven to be very effective, but its mechanism of action is still largely unknown \citep{Jakobs2019} \citep{aum2018deep} \citep{Lozano2019}. Studying the underlying therapeutic mechanisms of DBS first requires an in-depth understanding of the various activity patterns and differences in brain circuits associated with neurological disorders under physiological and pathological conditions. Therefore, combining computational modeling describing various types of neurons in the brain with clinical data is a rather promising research approach.

\subsection{Parkinson's disease(PD)}

PD is a complex multisystem neurodegenerative disease whose primary cause is thought to be the loss of dopaminergic cells in Substantia nigra pars compacta in the brain. The academic community initially considered it a movement disorder with three main symptoms: tremor, muscle rigidity, and bradykinesia  \citep{Postuma2015}. In addition to motor symptoms, PD also has many non-motor symptoms, including but not limited to autonomic dysfunction, sleep disturbances, and neuropsychiatric symptoms \citep{Hely2008}. PD is also the fastest-growing neurological disease in the world in terms of prevalence, disability, and deaths, with the GBD (Global Burden of Disease Study) estimating that the total number of people affected by the disease increased by 118$\%$ from 1990 to 2019, to more than 6 million worldwide  \citep{Dorsey2018}. With demographic changes (aging) and the side effects of industrialization (pollution), there has been a faster increase in PD cases \citep{Brundin2018}.

Various pathological features in PD are closely related to abnormal pathological changes in the basal ganglia (BG). The subthalamic nucleus(STN) in BG is the main target of DBS \citep{hu2017bibliometric}. Therefore, if we want to determine the specific mechanisms of DBS, it is important to first understand the physiological structure of the BG with its pathological changes in PD patients. Since the detailed mechanisms of DBS are difficult to investigate in humans, computational models are an effective tool to explore the mechanism of DBS and advance its development. 

In the next part, we will briefly introduce the main structure of BG and discuss the significant changes in the kinetic properties of BG-related loops in PD. 

\section{Basal ganglia(BG)}
\subsection{BG structure and functional masses}

The BG consists of a set of interconnected subcortical nuclei involved in various important brain functions - motor control \citep{turner2010basal}, decision making \citep{bogacz2007basal}, categorization  \citep{seger2008basal}, motor preparation \citep{jaeger1993primate}, probabilistic learning \citep{shohamy2008basal}, reinforcement learning \citep{Frank2005b}, reward signalling \citep{kawagoe1998expectation}, sequence learning \citep{lehericy2005distinct}, working memory \citep{mcnab2008prefrontal}, etc \citep{Brown1998} \citep{Nelson2014}. As shown in Figure 1, the several extensively connected subcortical nuclei are anatomically defined into striatum (STR), Globus pallidus (GP), subthalamic nucleus(STN), substantia nigra (SN), and ventral tegmental area (VTA). STR consists of caudate nucleus (CN), putamen (Put), and nucleus accumbens(Acb). GP can be divided into external pallidum (GPe) and internal pallidum(GPi). The SN is divided into pars compacta (SNc), pars reticulata (SNr), and pars lateralis (SNL) \citep{Aird2000}.

\begin{figure}[htbp]
\includegraphics[width=\linewidth]{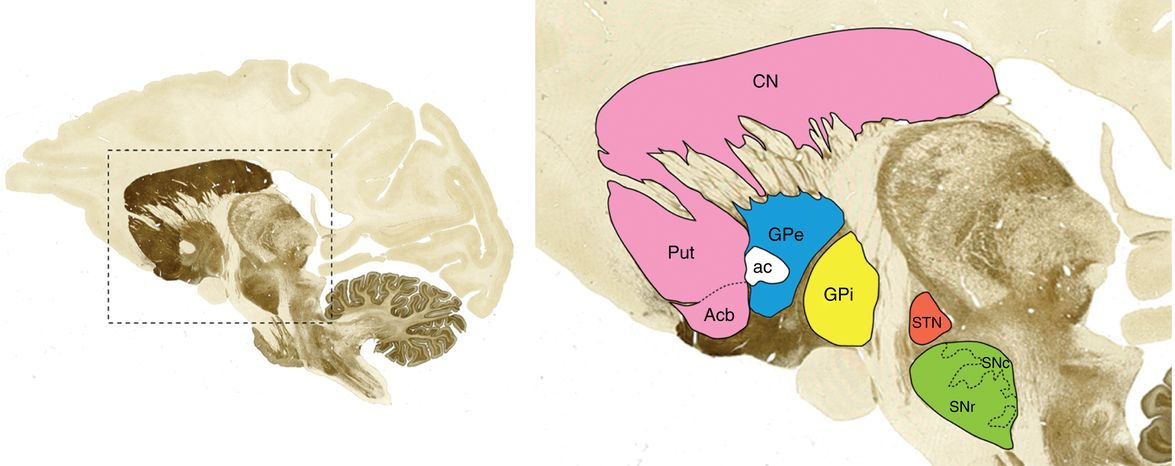}
\caption{\textbf{The location of basal ganglia along with its anatomy.(monkey brain) Referenced from\citep{Lanciego2012}} \\
STR consists mainly of spiny projection neurons. It receives input from the cerebral cortex, sensorimotor and motivational regions of the brainstem. It also produces either inhibitory or excitatory effects on the GP and SN, depending on the receptor type(D1 and D2-type) in the postsynaptic cell. \\
STN consists mainly of glutamatergic projection neurons. It receives input from cortical, thalamic, and brainstem structures and sends excitatory projections to the BG output nucleus and GP \citep{Nambu2002}.  \\
GPi consists mainly of GABA neurons and exerts strong inhibitory effects on neurons in the thalamus and brainstem \citep{Parent1999}. It receives STR input, inhibitory GABA input from GPe, and excitatory glutamatergic input from STN. The function of SNr is similar to that of GPi, and GPi/SNr is usually regarded as a single output structure of BG. \\
GPe consists mostly of large projection neurons. It is considered an important relay nucleus of the BG. It receives inhibitory input from the striatum and excitatory input from the subthalamus. It provides GABA inhibitory efferent connections to the input and output nuclei of BG \citep{Chan2005}. \\
SNc consists of large dopamine cells. It provides important regulatory signals to other BG nuclei and external structures (frontal cortex, septal area, amygdala, Etc.) and is an important component of the dopaminergic regulatory system in the BG \citep{Sulzer2005}.
}
\end{figure}

The BG comprises two principal input nuclei, STR and STN(also an important relay), and two principal output nuclei, SNr and GPi. GPe establishes connections between the output nuclei and other nuclei, acting as a relay. Dopaminergic neurons in SNc and VTA provide important regulatory signals to other BG nuclei.

\subsection{Synaptic connections and neural circuits in the BG}
Albin \citep{Albin1989} and DeLong \citep{DeLong1990}proposed two correlational theories in 1989 and 1990: The activity of subpopulations of STR projection neurons is differentially regulated by its afferent nerves. Different subpopulations of STR projection neurons may direct different aspects of motor control. Two important pathways of BG have been proposed: the Direct pathway and the Indirect pathway, known as the classical model of BG function. The classical model assumes that the output of BG is mediated by a direct pathway that promotes cortex activity (cortex-STR-GPi/SNr-thalamus-cortex) and an indirect pathway that inhibits cortex activity (cortex-STR-GPe-STN-GPi/SNr-thalamus-cortex). After the classical model was proposed, it was successively refined. The hyper-direct pathway was added  \citep{Nambu2002}.DeLong extended the concept of parallel loops proposed by Alexander et al. \citep{DeLong2010} \citep{Alexander1986}. Most researchers established the understanding that the indirect pathway inhibits movement while the direct pathway facilitates it. The selection mechanism for decision-making was further developed as the direct pathway is used for selection while the indirect pathway is used for control\citep{Gurney2001}.

\begin{figure}[htbp]
\includegraphics[width=\linewidth]{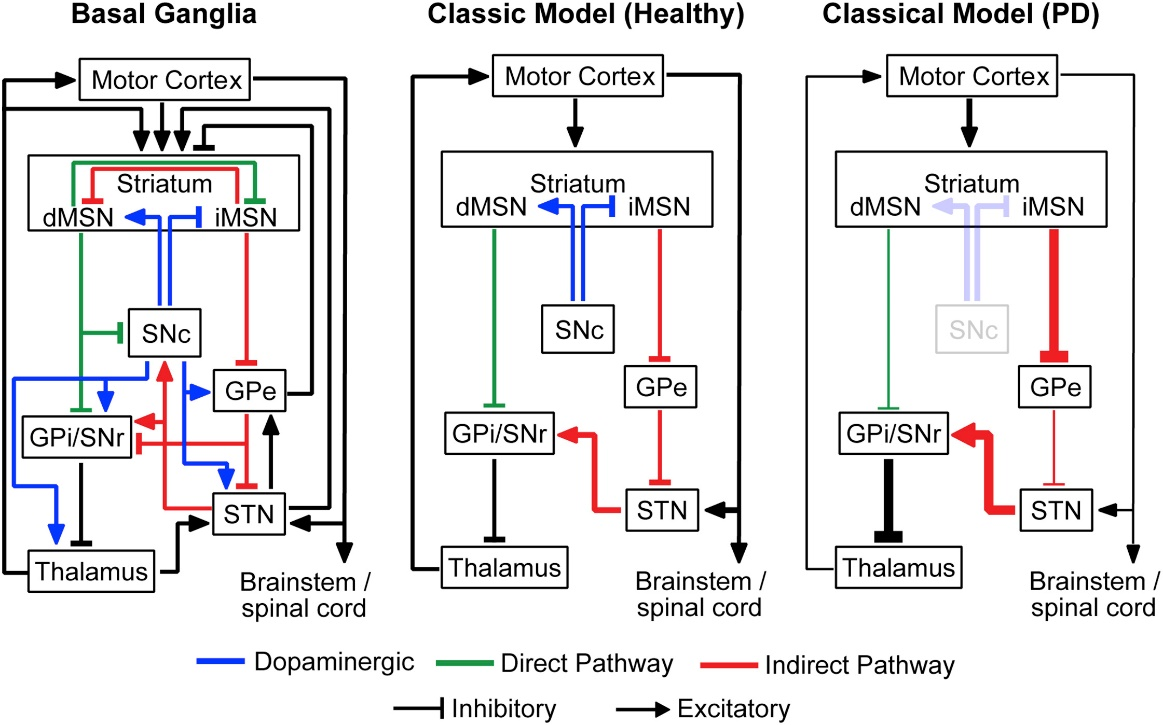}
\caption{\textbf{The classical model. }The left figure shows the original connection of BG. The middle figure is the schematic diagram of the classic model. The classical model describes the function of BG by simplifying the complex connections in BG into direct pathway and indirect pathway. The right figure shows the state of BG in PD. The apoptosis of SNC enhances the indirect pathway, resulting in dyskinesia. Referenced from \citep{McGregor2019} }
\end{figure}

\subsection{Pathology of BG in PD}
The degeneration of dopaminergic neurons of the SNc in BG of PD patients will decrease the dopamine levels in BG \citep{Jankovic2008}. It may change the BG function and cause an imbalance between direct pathway and indirect pathway. Studies have shown that the loss of BG dopamine in PD patients/primates is associated with sustained changes in firing rates and oscillatory synchrony 
between BG nuclei \citep{Boraud2005} \citep{Brown2007} \citep{Raz2000}. The direct pathway's effectiveness decreases while the indirect pathway's effectiveness increases while the firing rates of GPi and STN increase but the firing rate of GPe decreases. The imbalance in the firing activity of these neurons may cause the suppression of the motor system, resulting in various movement disorders \citep{Parker2018}. All of which have been clinically demonstrated \citep{Plenz1999}. 

When the firing rate of BG neurons changes, the firing pattern also changes. Abnormal local field potential (LFP) oscillatory activity occurs in the BG. The activities usually manifest as 13-30 Hz oscillations ($\beta$ oscillations), particularly pronounced in GPi, GPe, STN, and SNr \citep{Weiberger2006}. Symptoms of motor bradykinesia and rigidity in PD are thought to be closely related to $\beta$ oscillations in the BG \citep{Boraud2005}. Unlike the firing rate changes in the classical model of PD, the mechanism by which $\beta$ oscillations are generated remains unclear. Theoretically, any neuronal network containing negative feedback loops with delays can generate oscillatory activity patterns \citep{Ermentrout2001}. It has been suggested that $\beta$ oscillations may originate from a network of STN and GPe  \citep{Gillies2002}. Some computational models of PD-related neural oscillations have been proposed in the academic community \citep{Gillies2007} \citep{van2009} \citep{Sharott2005} \citep{Leblois2006} \citep{Holgado2010}.

\section{Mathematical theoretical model}
Many computational models have been developed to explain the state of BG-cortical loops in health and disease. It may also contribute to the understanding of the mechanism of DBS. The starting point for modeling is often based on explaining the neuronal mechanisms of BG function and PD symptoms. Since consensus on BG function is still pending, different scholars have chosen different theoretical approaches, resulting in computational models that differ in many aspects.

Making assumptions and simplifying relevant parameters in computational models is always important. Simplifying the necessary components to a minimum can improve the efficiency of computational models. To simulate the changes observed in the firing patterns of pathological BG, we can study neuronal activity at the neuronal level based on neurophysiological and neuroanatomical data. Thanks to the pioneering work of Alan Hodgkin and Andrew Huxley in explaining the generation and propagation of action potentials \citep{Hodgkin1989}, scholars have developed a well-established system of neurodynamic theory \citep{Gerstner2014} \citep{Devaney2010} \citep{Gerstner2002}. Most mathematical theoretical models have been developed on it. In other words, these models are based on the framework of biological theories rather than on specific patient conditions.

\subsection{Explore mechanisms}
\subsubsection{DBS restores the TC relay function}
To explore the effect of high-frequency stimulation of the STN on the STN-GPe loop and how their combined output changes the GPi output to the thalamus, Rubin and Terman developed a computational model in 2004 \citep{Rubin2004}. The model expanded from the previous STN-GPe model they developed in 2002 \citep{Terman2002}. GPi neurons and thalamocortical (TC) neurons were added to the model. Each neuron is constructed in a Hodgkin-Huxley type conductance-based model, represented as a set of nonlinear differential equations. Model neurons can replicate the firing phenomena of real neurons. TC cells of the model received a excitatory input train of spikes and inhibitory inputs from GPi. Figure 4 briefly represents the structure of the model neurons.

\begin{figure}[ht]
\includegraphics[width=\linewidth]{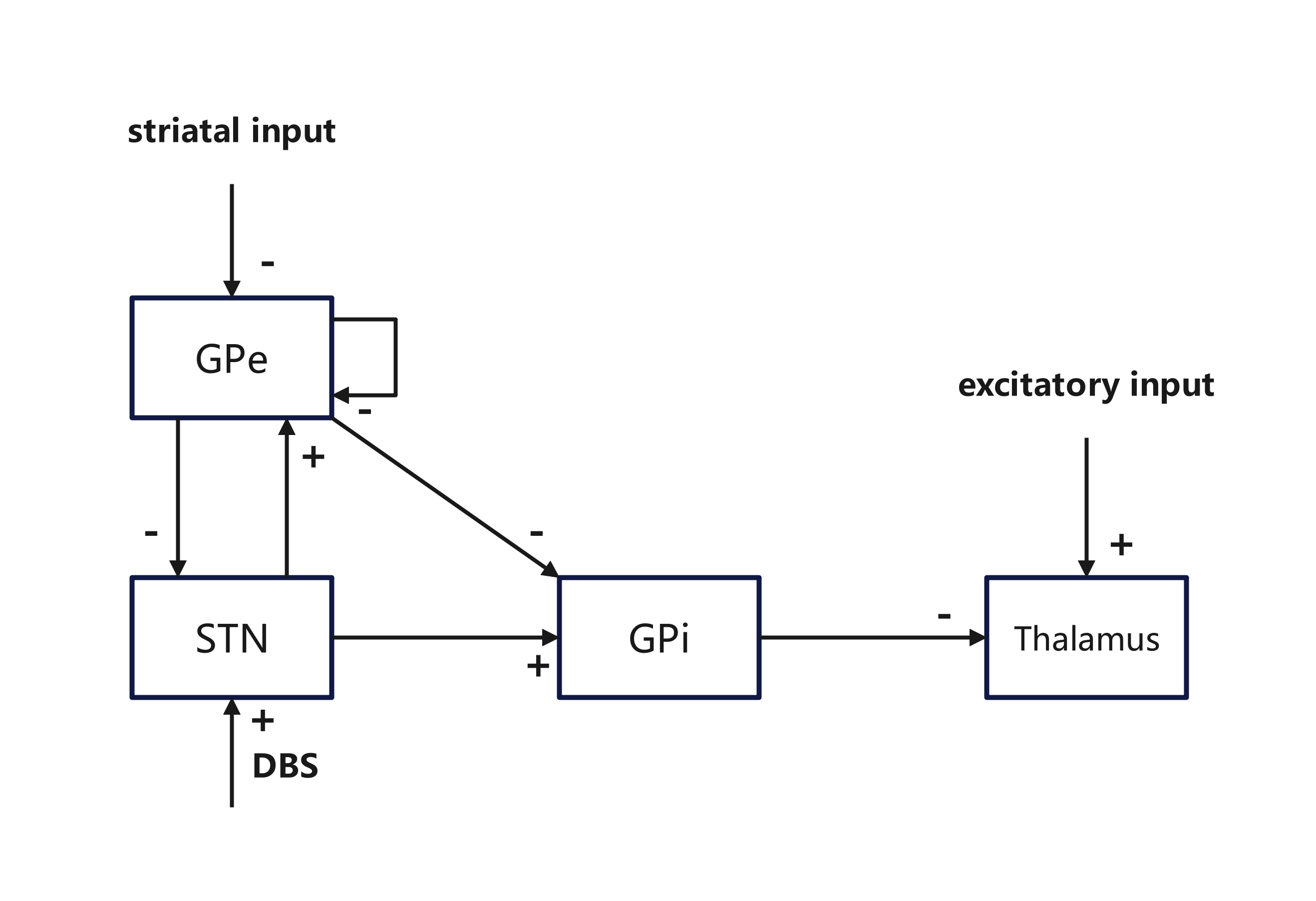}
\caption{\textbf{Structures in the Rubin and Terman model.} Arrows with - symbols indicate inhibitory synaptic connections and inputs, while arrows with + symbols indicate excitatory synaptic connections and inputs. DBS directly affects membrane current, which is applied to the STN by default in the model. Redraw from \citep{Rubin2004}}
\end{figure}

 The authors quantify the curative effect of DBS by using the error index, which represents the fidelity of the relay of TC cells to their excitatory inputs. The model has three states: normal, PD, and DBS. The output of GPi was irregular and uncorrelated in the normal state. Such output had a low inhibitory effect on thalamic cells, which can transmit excitatory sensorimotor input to cortical areas. The STR input and the GPi inhibition level were adjusted in PD state. The setting produced regular synchronized oscillatory activity in the STN-GPe network. The increased rhythmicity of STN and GPe firing resulted in rhythmic GPi firing. Experimental results suggested that the GPi neurons in the PD state induced $\beta$-frequency bursting discharges. The error index showed that the switch from irregular to bursting activity in BG impaired the fidelity of TC relay. TC cells were phase-inhibited and no longer completely transmitted motor sensory input of the cortex. In DBS state, the authors used high-frequency DBS at 167 Hz to stimulate STN neurons. The model implies that high-frequency stimulation induced a high-frequency firing rate of inhibitory GPi cells, which resulted in a strong but tonic inhibition. The tonic inhibition may have a much weaker effect on the TC’s ability to transmit motor sensory input than $\beta$ frequency inhibition. In other words, these changes improve TC cells' responsiveness to excitatory inputs. Thus, the authors concluded that DBS might effectively reduce motor symptoms in PD by eliminating the oscillatory nature of inhibitory inputs (e.g., GPi) to TC cells. 

 Rubin and Termen provide a better explanation of how STN DBS produces efficacy. Experiments have shown that high-frequency stimulation can promote the discharge of its stimulating target, or at least elicit effects consistent with its output enhancement\citep{hashimoto2003stimulation}\citep{hershey2003cortical}\citep{windels2000effects}\citep{garcia2005high}\citep{alavi2022excitatory}. If STN DBS enhances the GPi discharge so that it produces sustained, regular oscillations of GPi firing. Then according to the error index, the function of the thalamic relay is significantly improved. The idea provides the first mechanistic theory of DBS that regular spike trains in the BG affect the normal transmission of information in the brain. This mechanism hypothesis suggests that the regularization of GPi inhibitory input is more important than the level of inhibition itself. Thus, disrupting or eliminating this firing pattern may have a significant impact on the brain. 

The results of the Rubin-Termen model are very valuable, but it should be noted that the model still has many shortcomings due to the limited hardware and theoretical development at that time. The model can simulate the burst-like firing pattern, but the frequency dependent effects of STN-DBS on PD cannot be reproduced. It does not think about 3D orientation of different neural nuclei and location of stimulation electrodes. So et al. 2012 revised this model\citep{so2012relative}, but when the DBS frequency was lower as 20 Hz, the stimulation became effective. Thibeault et al. 2013 replaced the neuron models with Izhikevich hybrid neurons for less computational cost in large-scale biophysical models\citep{thibeault2013using}. Karthik et al. 2016 again revised the model, to better simulate experiment results of 6-OHDA lesioned model\citep{kumaravelu2016biophysical}.

\subsubsection{DBS activates soma and axon separately}
Close to the time when the model was proposed by Rubin et al., McIntyre and M. Grill used a finite-element model and a multicompartment cable model to study DBS \citep{McIntyre2004}. They focused on individual neuronal responses to DBS. They studied electrophysiological properties of the thalamic cortex, resting neurons and firing neurons with DBS administration. They concluded that the neuronal response to DBS depends mainly on the position and orientation of the axon relative to the electrode and the stimulation parameters. 

\begin{figure}[htbp]
\includegraphics[width=\linewidth]{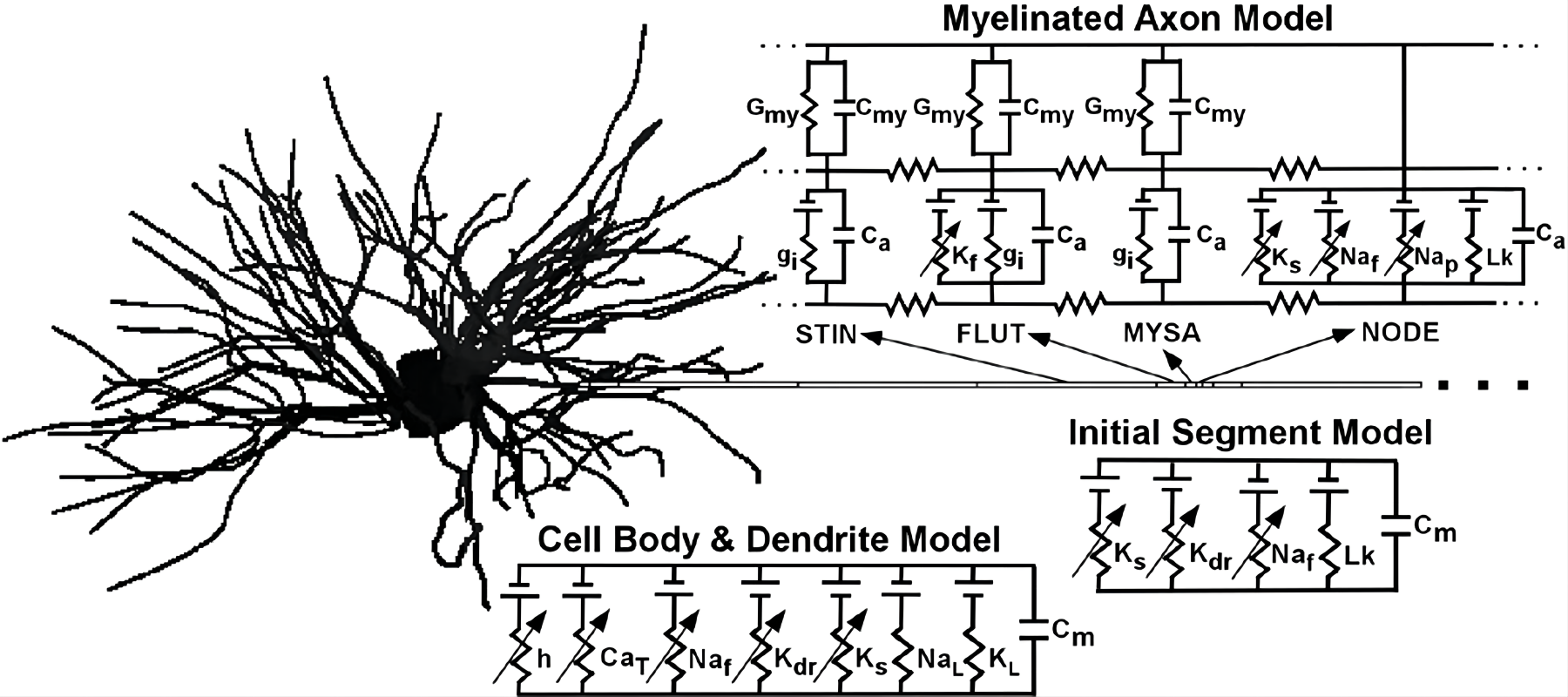}
\caption{\textbf{Cable model of McIntyre and M. Grill.} Model consisted of a 3D branching dendritic tree, multi-compartment soma, and initial segment, and a myelinated axon. Referenced from \citep{McIntyre2004}}
\end{figure}

In McIntyre's model, direct activation of TC relay neurons by subthreshold stimulation is inhibited by intrinsic firing rate (spiking firing or bursting firing) activity in the stimulus train mediated by activation of presynaptic terminals. Suprathreshold stimulation causes inhibition of intrinsic firing rate in the soma but produces efferent output in axons at the stimulation frequency. In a nutshell, DBS could inhibit activity in the soma while stimulating axons. They considered the independent activation of efferent axons. In their results, axons and cell bodies are independent. The model showed that DBS overrides oscillatory pathological activity and replaces it with more regularized neuronal firing patterns. The concept was called "Informational Lesion" by Grill et al.\citep{grill2004deep}. The model provided a new basis for the hypothesis of a stimulus-induced regulation of cellular activity as a therapeutic mechanism for DBS. 

\subsubsection{Gpi burst activity in DBS}

In 2010, Hahn and McIntyre built a point-neuron model of network dynamics to study the rate and pattern of subthalamopallidal network activity in DBS\citep{Hahn2010}. They built the computational model of the hypothalamus-pallidum base on microelectrode recordings from the basal ganglia (BG) of non-human primates in order to better compare DBS network model predictions and the underlying neurophysiology. Each neuron is a current containing Hodgkin-Huxley type neuron. After training the model to fit in vivo recordings from PD monkeys, the model was used to assess the effects of STN-DBS. The model reproduced the result that reduced GPi activity during DBS in PD monkeys. 

\begin{figure}[htbp]
\includegraphics[width=\linewidth]{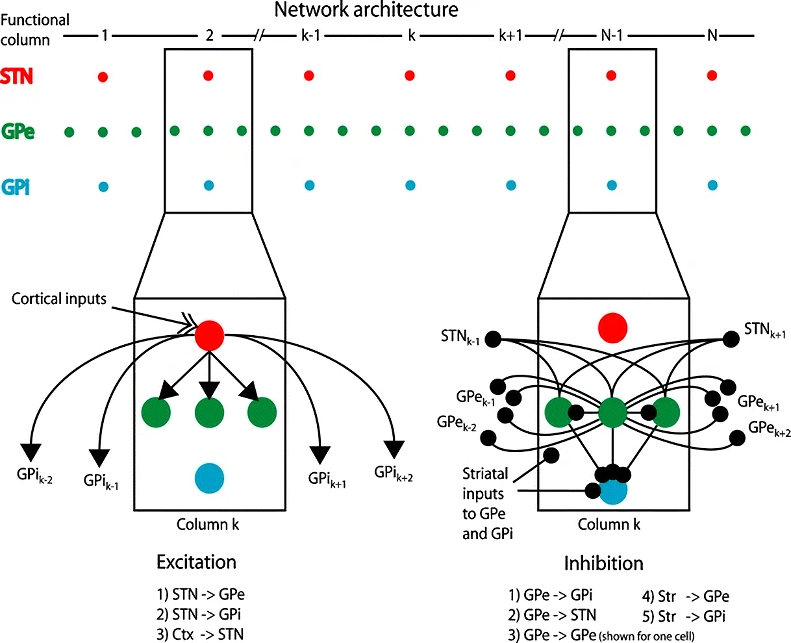}
\caption{\textbf{Structure of the hypothalamic pallidum network of McIntyre et al.}  Excitatory connections are on the left side and inhibitory connections are on the right side. The external inputs to the model are excitatory cortical inputs to the STN and inhibitory STR inputs to the GPe and GPi. Referenced from \citep{Hahn2010}}
\end{figure}

In their analysis of GPi burst activity, they proposed three main predictions: \textbf{a)} The volume of the DBS-activated STN is positively related to GPi burst activity. \textbf{b)} DBS stimulation frequency also reduces GPi burst activity and builds its best effect at clinical frequencies. \textbf{c)} Destruction of STN neurons could also produce the same effect as DBS treatment.

\subsubsection{Resonant frequency determines the optimal DBS frequency}
Although the models mentioned above are very representative, it contains a relatively small number of neurons. It assumes that the STN is uniformly activated for analysis. These settings may result in the model losing some critical information in the simulation of real situations. Therefore, some scholars have scaled the neuronal model to study the heterogeneous and macroscopic network effects of STN high-frequency stimuli when extended to the whole network \citep{Modolo2007} \citep{Humphries2012} \citep{McIntyre2010}. These models are usually large, containing hundreds of neurons. 

Humphries and Gurney modified the rat basal ganglia network model they developed in 2006\citep{humphries2006physiologically} to study DBS\citep{Humphries2012}. They set the intensity of stimulation to neurons as related to the distance between that neuron and the stimulating electrode in their model. Humphries used the model to simulate the heterogeneous responses in STN high-frequency stimulation and the heterogeneity of the changes in spiking and bursting firing of GPi neurons. They replicated the changes recorded from primate GPi during STN high-frequency stimulation. These recorded changes suggested that the model predicts a mixed response in the neural network that is consistent with STN-DBS treatment. Finally, Humphries et al. concluded that STN-DBS resets a natural output balance to the GPi. These large network effect models also provided a possible explanation for why stimulation frequencies above 100 Hz are clinically effective stimulation frequencies: the optimal DBS frequency is the resonant frequency of the neuronal network.

\begin{figure}[htbp]
\includegraphics[width=\linewidth]{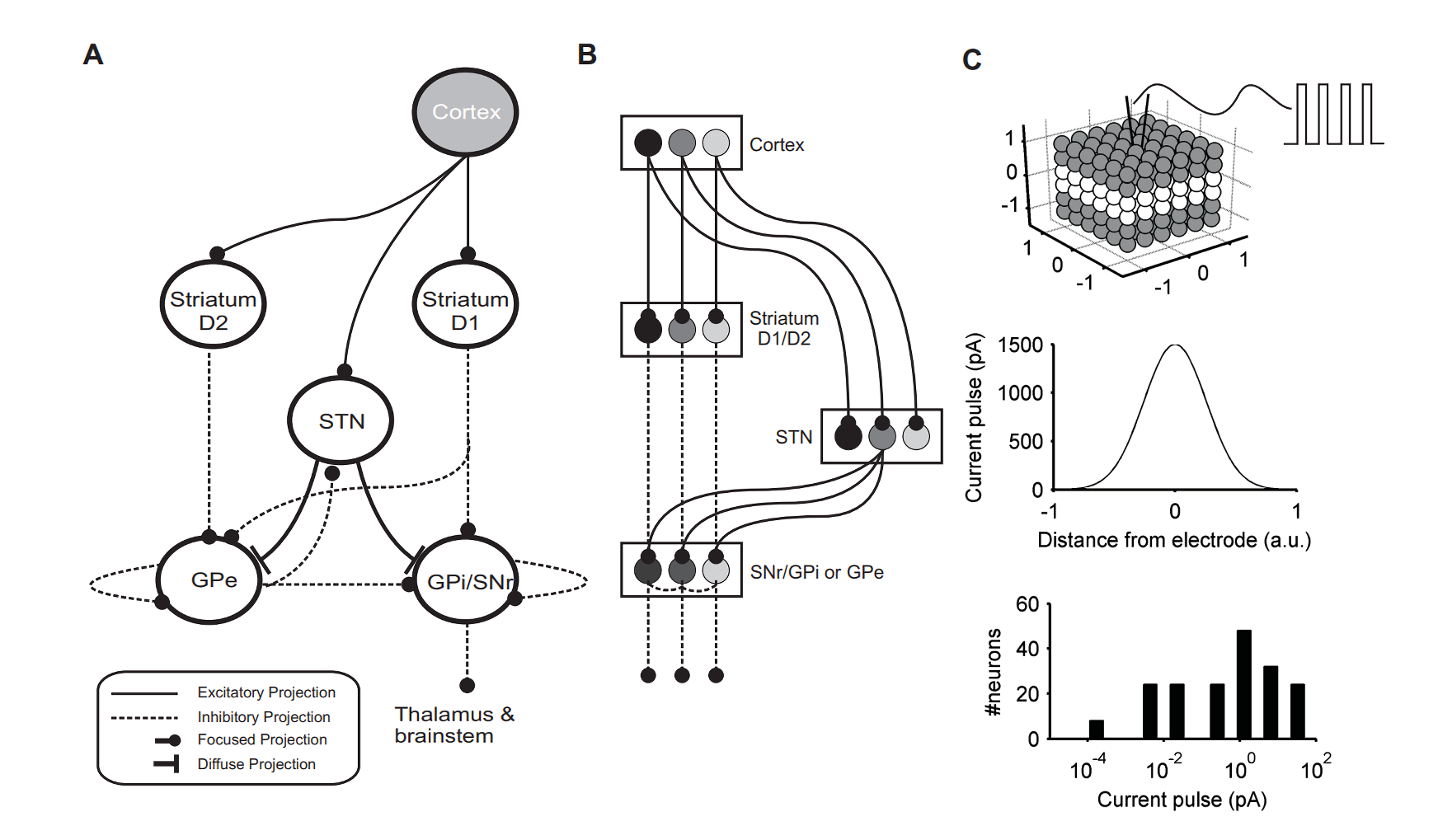}
\caption{\textbf{Humphries and Gurney's network model. } (A) BG macrocircuit. The GABAergic cortical input reaches the STR and the glutamatergic thalamic nucleus (STN). (B) The main circuit can be decomposed into two copies of the eccentric surround sound network. (C) Deep brain stimuli are modeled as propagating according to network effects. Referenced from \citep{Humphries2012}}
\end{figure}

 Gradinaru et al. innovatively used optogenetic techniques  \citep{Gradinaru2009} to develop an optogenetic DBS model. They proposed a perspective on the effect of high-frequency pulses on DBS in the STN by driving the spiking activity of STN neurons expressing retinoid channels with 130-Hz light pulses. In their model, 130-Hz STN stimulation did not provide relief from behavioral deficits in PD rats. Their result challenged the conclusion drawn from the model above that regulated high-frequency STN spiking activity is critical for the effectiveness of DBS treatment. Humphries compared the optogenetic DBS model with his construction of an electrical DBS model \citep{Humphries2012} \citep{MD2018}. They found that the DBS in the optogenetic DBS model and the electrical DBS model differ in their effects on the rate, regularity, and correlation of BG output.

From these models, we can learn that DBS can alleviate motor symptoms in PD, but not necessarily by restoring the diseased neurons to their normal state to achieve the effect.

\subsection{Testing different stimulation targets and stimulation patterns}
Therapeutic DBS provides continuous high-frequency cyclic stimulation of brain tissue in the target site of stimulation. The almost "brutal" treatment provides good therapeutic results for patients while causing many undesirable side effects \citep{zarzycki2020}. Changing the stimulation target and designing new stimulation patterns may improve these side effects and increase efficacy. These improvements can improve the quality of life of patients. For example, new low-amplitude stimulation patterns may extend the life of the stimulator battery and reduce the DBS current spillover beyond the target site. However, the exploration of new modalities and targets can be met with greater resistance in the clinic due to ethical constraints. Computational models offer the possibility of many new stimulation targets and stimulation patterns. Many computational models have been proposed for testing new stimulation targets and modalities.

\subsubsection{Different targets of DBS} 
Pirini et al. used a modified version of the Rubin-Terman model to investigate the performance of BG networks in different stimulus targets  \citep{Pirini2009}. SMC sensorimotor cortex was added to the model, and the simple constant currents represent GPe, GPi, and STR inputs. DBS was modeled as a train of rectangular positive current pulses injected directly into cells belonging to the target site. Pirini applied DBS at 30 Hz, 120 Hz, and 180 Hz to three stimulation targets, STN, GPe, and GPi. 

\begin{figure}[htbp]
\includegraphics[width=\linewidth]{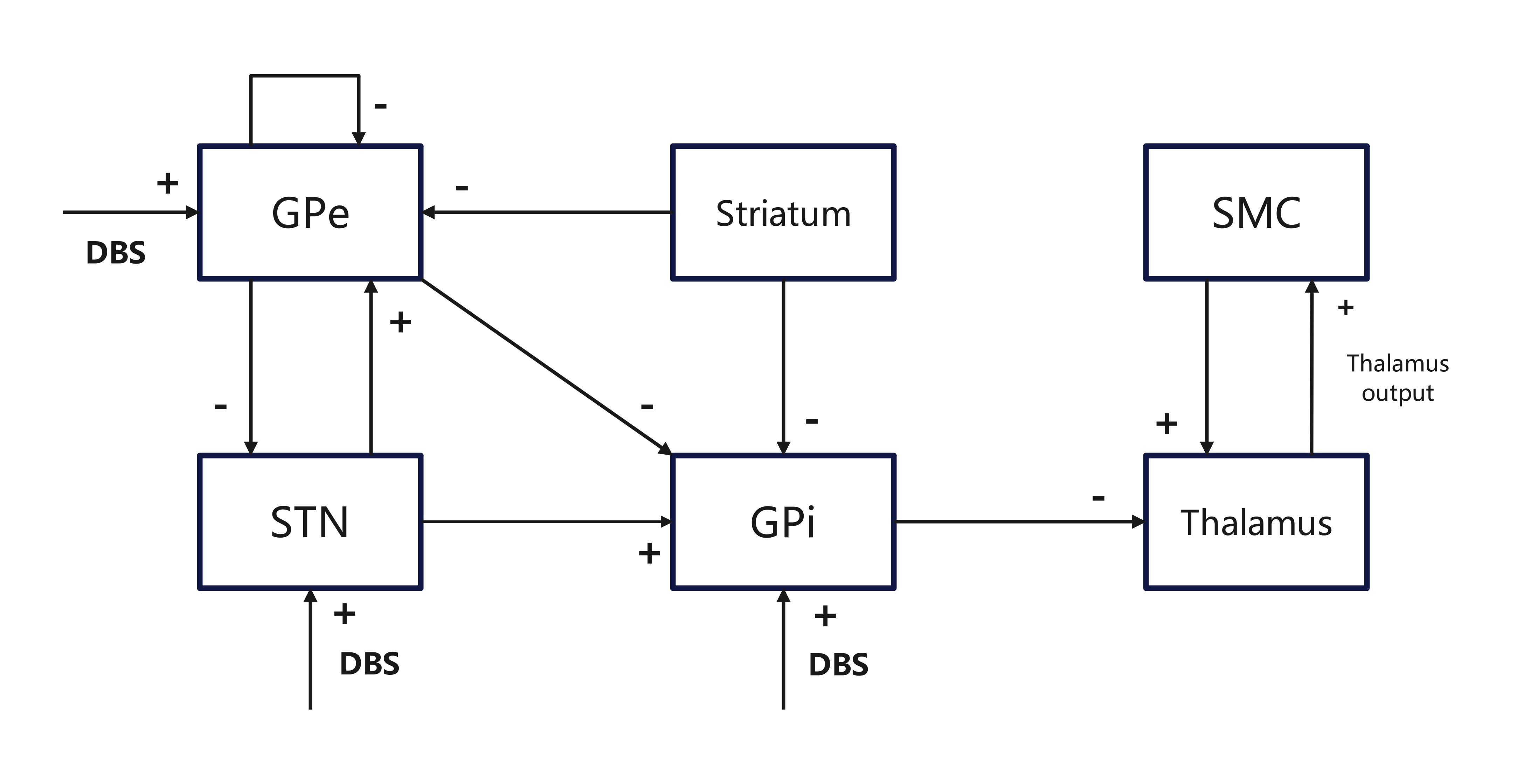}
\caption{\textbf{The Rubin-Terman model modified by Pirini et al. }DBS can act on either structure in Fig. Redraw from \citep{Pirini2009}}
\end{figure}

Experimental results showed that low-frequency GPe-DBS and GPi-DBS (30 and 60 Hz) were insufficient to drive STR to cause benign effects on TC relay function. STN low-frequency stimulation only partially restored TC relay function. It could partially explain the theory that lower than low-frequency DBS worsens PD symptoms \citep{Perlmutter2005}. High-frequency STN-DBS restores TC relay function, whereas DBS in GPe and GPi over-activate and inhibit it. These simulation results are consistent with experimental results on DBS network effects achieved in human subjects and monkeys by microdialysis and extracellular recording procedures \citep{Stefani2005} \citep{Stefani2006}.

\subsubsection{Large network model}
To investigate the effect of new stimulation patterns on the BG, Kumar and colleagues developed a large network computational model consisting of 3000 neurons \citep{Arvind2011}. They stimulated the STN with two different stimulation patterns. Then they observed and analyzed the role played by these two patterns on the network model. In the periodic stimulation mode, Kumar found that neither excitatory nor inhibitory STN stimulation inputs were critical to making a $\beta$-oscillation reduction. In contrast, the excitatory and inhibitory inputs of periodic high-frequency stimulation above 100 Hz reduce $\beta$-oscillation generation. Kumar then designed a non-periodic stimulation pattern such that the pulse period of DBS was $\gamma \Delta t$. ($\Delta t$ is the minimum interval between pulses, fluctuating between 0 and 15 ms. $\gamma$ is a randomly chosen integer value between 0 and 3) They found that the stimulation could suppress pathological oscillations more effectively than periodic stimulation.

\subsubsection{Test different stimulation patterns }
Brocker and M. Grill et al. used the Rubin-Terman model to test new stimulation modalities. The three off-cycle stimulation modalities they designed showed promising properties \citep{Brocker2013}, including but not limited to having lower battery consumption, lower stimulation frequency, and better reduction of dyskinesia. The thalamic DBS mode with brief pauses was subsequently tested. It was found that stimulation pauses reduced the efficacy of open-loop DBS to suppress tremors, suggesting that thalamic DBS may reduce tremors by masking pathological burst discharges of PD in the BG \citep{Swan2016}. 

\begin{figure}[htbp]
\includegraphics[width=\linewidth]{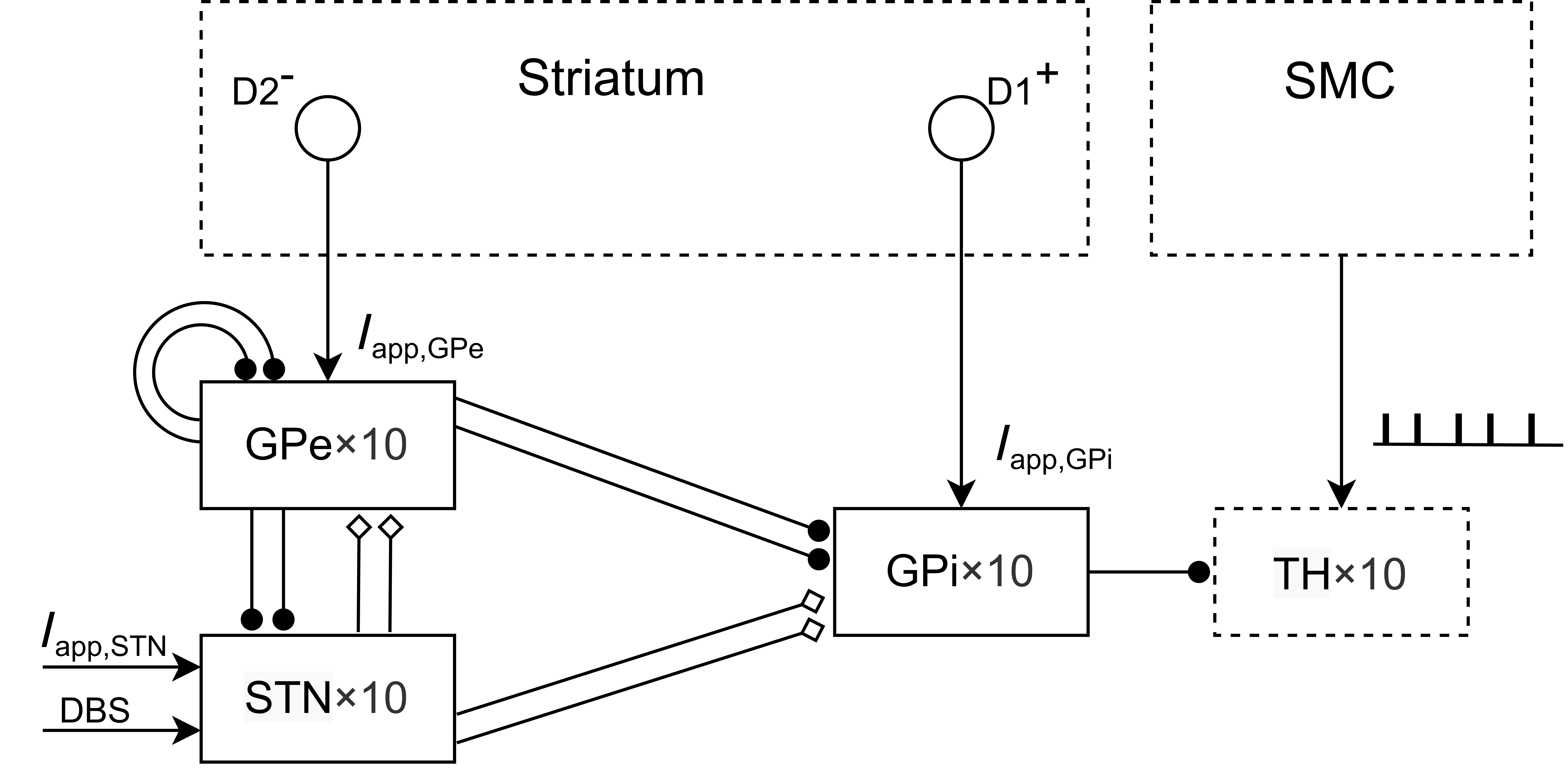}
\caption{\textbf{The model of Brocker and M. Grill et al.} Diamonds ($\diamond$) and loops ($\bullet$) indicate excitatory and inhibitory synaptic actions, respectively. Redraw from \citep{Brocker2017}}
\end{figure}

Brocker and M. Grill used a genetic algorithm to apply these different stimulation modes to the model through multiple iterations. They assessed the effect of each stimulus pattern on the network signal transmission to select the best stimulus pattern for the lower frequency  \citep{Brocker2017}. After that, Cassar and M. Grill proposed and evaluated several new modifications to the genetic algorithm. The proposed genetic algorithm modifications significantly reduced the number of iterations required for convergence and identified good stimulus patterns \citep{Cassar2017}.

\subsubsection{Coordinated Reset}

As mentioned above, abnormal neuronal synchronization is one of the hallmarks of PD, in addition to $\beta$ oscillations. Normal DBS stimulation is quickly followed by the reappearance of PD symptoms after cessation. To address these two points, Peter A Tass developed a new simulation method using a mathematical-theoretical model of a coupled phase oscillator that simulates synchronized neurons. The stimulation pattern leads to pathologically sustained desynchronization and "forgetting" of synaptic connections, resulting in a good therapeutic effect for the patient. Tass called it "coordinated reset" (CR) \citep{Tass2003}. During CR stimulation, phase-shifted stimuli are delivered to multiple stimulation sites to counteract the synchronization of neurons. Compared to other stimulation modalities, CR does not require precisely timed stimulus delivery, has a better therapeutic effect on the patient, and can reduce side effects. Tass and colleagues then conducted an in-depth study of spike-time-dependent plasticity (STDP) in the nervous system. They found that an uncorrelated firing state and pathological synchrony state can coexist due to the STDP, making it possible for DBS's therapeutic effects to persist even after stimulation is stopped \citep{Tass2006}. After developing and testing the theory in a computational model, the method was tested in a non-human primate model of PD \citep{Tass2012_b} and PD patients \citep{Adamchic2014}. The results showed that DBS efficacy using the theory persisted for many days after stimulation was terminated. Later Tass et al. extended CR stimulation to the more general field of neuromodulation and found that sensory stimulation and auditory stimulation also improved pathological neuronal synchronization phenomena in PD patients \citep{Popovych2012} \citep{Tass2012_a}.

\subsection{Closed-loop DBS model}
In clinical practice, most current DBS systems are open-loop, and the physician's experience in the clinic mostly determines parameter adjustments. Therefore, DBS systems cannot automatically optimize the stimulation settings to provide different stimulation patterns to different patients \citep{Ramirezzamora2017}. Compared to open-loop DBS, closed-loop DBS is a more intelligent stimulation modality. Its stimulation parameters vary with the physiological variables monitored by the electrodes and have a more flexible stimulation pattern. Closed-loop algorithms that adjust stimulation parameters based on patient physiology can help to identify the optimal settings more systematically and efficiently. Computational models can act as virtual patients and are ideal platforms for the first testing and development of adaptive closed-loop algorithms before applying them to patients or even animals. Closed-loop control of DBS is more effective in treating PD than open-loop configurations \citep{Rosin2011}. However, the device and patient situation limit closed-loop algorithms and generally need to satisfy the following conditions (expanded by the guidelines proposed by Feng et al.  \citep{Feng2007}): (1) The algorithm must directly utilize clinical observations that can be measured quickly. (2) The algorithm must validate or falsify the hypothesis in a computational model, in vitro or in vivo. (3) The algorithm needs global search capability to fully utilize the DBS parameter space. (4) The algorithm should be able to operate with different types of DBS stimulation patterns (e.g., periodic and non-periodic stimulation) (5) The algorithm should be able to have good advantages over open-loop mechanical stimulation. (6) The algorithm should have good robustness and be less susceptible to dramatic fluctuations from changes in clinical observables.

A key step in developing such a system is to identify signal characteristics or "biomarkers" that may quantify the clinical state. In the closed-loop control of DBS in PD, one of the most promising signal features is the level of $\beta$-band (13-30 Hz) oscillatory activity in the STN and cortex-BG networks  \citep{Chen2006} \citep{Kuhn2008} \citep{AA2009}. In addition to the local field potential (LFP), signal features such as microelectrode recordings (MER), imaging data, and external sensors are worth exploring. Moreover, there are various methods to adjust stimulation parameters, such as recursive extended least square method \citep{zhu2021adaptive}, machine learning \citep{Merk2022}, proportional-integral controllers \citep{Fleming2019}, fractional-order controllers \citep{Coronel2020}, and genetic algorithms \citep{Cassar2017}.

\subsubsection{Phase response curves in closed-loop DBS}
Holt and Netoff used the model of DBS developed by Hahn and McIntyre to investigate closed-loop DBS\citep{holt2014origins}. They created a mean-field model of the closed-loop system, consisting of discrete time models and their Z-transforms. The simplified mean-field model faithfully reproduces the features of Hahn and McIntyre's model. The authors proposed that the phase response curve of the $\beta$ oscillation’s response to subthreshold DBS-like pulses is a physiological measure that can be used to predict optimal stimulus frequencies and developed a new method for determining phase response curves from population data. The Fourier coefficients of spike times at 34Hz around low-frequency stimulation were used to estimate the phase of the population and instantaneous amplitude. Curve fitting was then performed on the phase change as a function of the stimulus phase to estimate the phase response curve. The authors showed that 
the phase response curve can be used to predict the effect of stimulation frequency on the oscillation amplitude. In 2016, they further studied\citep{holt2016phasic}, showing that the PRC can be used to predict the effects of phasic stimulation on the amplitude of the 34 Hz oscillation, and proposed a phasic burst stimulation protocol optimized using the PRC. The power of pathological 50 Hz oscillations was reduced by nearly 34$\%$ with phasic burst stimulation. 

\subsubsection{Nonlinear predictive control and radial basis function guide}
Wang Jiang et al. used the basal ganglion network model developed by Rosa et al. to test the closed-loop stimulation method\citep{su2018nonlinear}. They used a nonlinear predictive control algorithm to achieve online tuning of stimulus parameters and then tried to understand the difference between conventional DBS and closed-loop stimulation from the distribution of interspike intervals of BG neurons. The LFP signal of the GPi was used as a feedback signal. 
The authors assumed that the relationship between the input stimulus signal and the output LFP signal was nonlinear, so a second-order autoregressive Volterra model was used to identify the input-output relationship. By comparing the inter-spike-intervals distributions of various stimulation frameworks, the authors found that their proposed closed-loop system was more therapeutically effective than other stimulation modalities.
In 2012, Wang Jiang et al. collaborated with M. Grill to test the new closed-loop stimulation algorithm in a biophysically-based model of the cortex-basal ganglia-thalamus network developed by Kumaravelu et al\citep{zhu2021adaptive}\citep{kumaravelu2016biophysical}. Guided by the radial basis function network, both the p and pi controllers effectively suppressed the generation of $\beta$ oscillations in the GPi nucleus cluster. These two attempts partially demonstrated the feasibility of $\beta$  power as a feedback signal in DBS.

\section{Clinical Predictive Models}
In contrast to theoretical mathematical models, patient-based computational models are mostly used in the clinical field. These clinical prediction models help clinicians to predict patient-specific outcomes of DBS and to understand the variability of outcomes across patients \citep{McIntyre2007}. They allow a non-invasive quantitative description of 130the theoretical responses of various neural elements to different stimulus settings. Since the success of DBS surgery depends largely on the accurate placement of electrodes in the brain, 3D brain models generated from imaging data and patient signs can effectively assist the surgeon in the procedure and thus reduce patient risk. Also, because these models simulate the anatomy of the electrodes and surrounding tissue, fitting the models to clinical data allows exploration of the volume of tissue activation (VTA) produced by stimulation and helps to adjust stimulation parameters. Finally, new electrode lead designs can be tested in computational models first.

\subsection{Optimize DBS surgery}

Presently, the electrode placement and stimulation parameter adjustment of DBS are mostly based on the experience of doctors. The calculation model based on patient imaging data helps doctors understand the situation of patients and help doctors accumulate experience. MiocinovicS and colleagues developed a modeling tool, Cicerone. It is based on Microsoft Windows for clinicians or researchers to establish DBS-related models more conveniently and effectively. It can predict VTA to help doctors visualize the impact of various stimulus parameters on surrounding anatomical structures \citep{Butson2007} \citep{Miocinovic2007}. Cicerone can also visualize MRI, CT scan, 3D brain atlas, MER data, and the interaction between DBS electrode and VTA in three-dimensional, showing patients' specific situations more clearly in the eyes of doctors and researchers. The software or other similar systems \citep{Finnis2003} \citep{DHaese2005} is an example of how to use the DBS model to enhance DBS surgery in the future. Preoperative planning allows the definition of stereotactic anatomical targets and trajectories. During operation, stereotactic micro drive coordinates and MER data can be input to realize 3D real-time interactive visualization of electrode position relative to peripheral neuroanatomy and neurophysiology. In addition, neurosurgeons can combine anatomical (such as MRI/CT/3D brain atlas), neurophysiological (such as MER), and electrical (such as DBS VTA) data to optimize the preoperative planning of DBS electrodes before 130permanent implantation.

There are many short-term adverse effects after DBS \citep{hu2017short}, and more accurate postoperative modeling may help physicians understand the causes of these adverse outcomes and help patients alleviate their symptoms. LEAD-DBS, a MATLAB toolbox, is a similar and relatively novel system with similar functionality to Cicerone \citep{Horn2015}. Horn developed the automated DBS electrode reconstruction algorithm in 2015 after analyzing 50 patients who underwent DBS procedures. The toolbox focuses on semi-automatic DBS electrode reconstruction based on postoperative MR or CT data and can visualize DBS lead trajectories and mapping data. The toolbox was used in various clinical cases  \citep{Vasileios2020} \citep{Aaaa2022} \citep{Yu2022} or biological models such as rats \citep{Andree2022}, all with good results.

\subsection{The model customized for patients.}
A limitation of DBS optimization is the need to find clinically effective stimulus parameter settings within the stimuli window that do not cause patient damage, such as efficacious windows that cause side effects. The limitation is a tedious and time-consuming process. Patient-specific DBS computational models based on imaging data and volume conduction models can assist in the process of DBS program control. Quantitative theoretical predictions can be used to define effective stimulation parameter settings. Such models are tailored to the patient and maximize stimulation to defined target areas while minimizing stimulus propagation to non-target areas.

Patient-tailored model contains two important components: imaging data and volume of tissue activation (VTA); imaging data have been described above. Since the effects of electric currents on brain tissue are quite complex and there is still much uncertainty in the academic community about such effects, acquiring VTA requires a combination of means. Finite element models can be used to predict how electric fields penetrate brain tissue  \citep{Cameron2001}, so generating finite element models from imaging data and pairing them with physiologically realistic neuronal models can accurately predict VTA \citep{Butson2005}.

Although DBS has been increasingly used in the clinic, researchers and physicians still know little about the neural response to applied voltage distribution. They have difficulty relating the modulation of different brain pathways to clinical prognosis. In other words, we want to understand how the specific axonal pathways activated in DBS affect the treatment outcome. To address these issues, Gunalan and McIntyre et al. proposed the pathway activation model (PAM) \citep{Gunalan2017} based on previous work \citep{Lujan2012} \citep{Lujan2013}. The PAM considers many factors representing the axonal response in DBS, which are more comprehensive than some diffusion-weighted (DWI)-based fibre bundle, imaging models. These factors include
\begin{enumerate}
\item the electrode configuration,
\item the shape, duration, and frequency of the applied stimuli,
\item the electrical conduction properties of the brain tissue medium,
\item the geometry and trajectory of the axons,
\item the membrane biophysics of the axons.
\end{enumerate}

Gunalan and McIntyre et al. attempted to simulate changes in the hyper-direct pathway in DBS with the model. They validated the model in a 67-year-old male patient who had PD for about 11 years and found that their proposed model predictions were in good agreement with the clinical hypothesis. That is, the hyper-direct pathway is directly activated during therapeutic STN-DBS, in agreement with the theory of Li et al. 

These patient-specific field-cable pathway activation models are very accurate for quantifying cellular responses to applied electric fields, but like other models, they have several limitations and drawbacks. For example, PAM cannot quantify the network-level modulation of DBS, and its Multi-chamber cable model of axons ignores some of the complex branching patterns of real axons. In addition, PAM is so technically demanding to implement that its use in clinical studies is limited. Gunalan and McIntyre later developed the DF-Howell algorithm to improve the deficiency  \citep{Howell2019}.

\subsection{Driving new DBS electrode designs}

The therapeutic effect of DBS and the amount of brain tissue activation are influenced by the physical properties of the electrodes and the stimulation parameters \citep{Mcintyre2002}. Most of the DBS electrodes currently used surgically are cylindrical with 4-8 contacts around the lead, equally spaced along the length of the lead. Computational models can be used to investigate whether different lead designs lead to differences in electric field distribution \citep{Fabiola2018} and help analyze the effects caused by them. Partitioning the stacked cylindrical electrodes into three-dimensional electrode arrays along and around the DBS electrodes may facilitate more spatially targeted stimulation and recording from deep brain structures. Computational modeling studies have shown that subdividing the circumference of each cylindrical electrode into four rectangular electrodes \citep{Brinda2021} or four oval electrodes \citep{Vitek2019} \citep{Pena2017} \citep{Lauri2017} can improve the ability of the lead to control, move, and affect areas of stimulated tissue within it. Connolly et al. applied the design to the clinical setting by designing new electrodes that segment the ring around the electrodes and found experimentally that such electrodes can control current more effectively \citep{Allison2016}. Zitella and colleagues tested two computational models and found that current control using segmented electrodes may help focus stimulation when the electrodes are slightly off-target or targeted to brain regions with complex geometry  \citep{Zitella2013}. The DBS electrodes with radially distributed electrodes have the potential to improve clinical outcomes by more selectively targeting pathways and networks within the brain  \citep{Teplitzky2016}. 

Directional DBS electrode designs are now commonly used in clinical practice, with newer directional electrodes segmenting the cylindrical band into three separate electrode contacts. These segmented electrodes have shown promise in redirecting stimulation to the treatment area and away from areas known to produce side effects. Thus, directional DBS electrodes can increase the therapeutic window for clinical stimulation by decreasing the efficacy and side effect threshold. Recently Gilbert et al. compared different approaches for representing targeted DBS electrodes in a finite element volume conductor (VC) model \citep{Frankemolle2021}. They evaluated 15 different variants of the DBSVC model and quantified how their differences affected estimates of the spatial extent of DBS axon activation. The result supposes that different DBS-VC models yielded significant differences in their voltage distributions and axon thresholds. For example, more complex VC models take 2-3 times longer to mesh, construct, and solve for the DBS voltage distribution compared to simpler VC models.

\section{Discussion}
After more than three decades of development, DBS has become an established intervention for PD and is a well-established treatment recognized by the medical community. However, the exact mechanism of DBS and the optimal site and stimulation parameters remain unresolved, so most treatments are still largely based on trial and error. During the first two decades of DBS development, the developing of stimulation sites and devices for DBS was slow. In any case, in the last decade, computational models have contributed to some extent to the emergence of multiple new stimulation methods and devices that have effectively advanced DBS technology. DBS models have extensively explored major phenomena in PD pathologies, such as beta oscillations, dopamine depletion, and dysfunction of direct and indirect pathways. This review reviews several representative computational models of DBS. Their purposes include, but are not limited to, exploring the mechanisms of DBS, investigating better stimulation targets, optimizing treatment options for clinical patients, and designing new DBS stimulation devices. Although the work to date does not provide an authoritative explanation of the exact mechanisms of DBS concerning the aetiology of PD, the variety of computational models proposed by scholars targeting specific PD phenomena and DBS effects provides a wealth of tools for researchers and physicians.

Each model has its limitations. For example, the basic connectivity within the BG network is much more complex than the model, the model does not take into account the differences between neurons, and the effects of DBS electrical stimulation are oversimplified in the model. Despite these shortcomings, the DBS computational model still provides great help when studying the pathophysiological mechanisms of PD and the potential mechanisms of DBS action. After comparing the models in the review, it is easy to see that mathematical theoretical models attempt to simplify the various influences in DBS in order to analyze them mathematically and thus understand the nature of DBS. In contrast, clinical prediction models tend to consider multiple influences to provide a better analysis of the patient's condition and to obtain a better treatment outcome, making their models very complex and difficult to parse. Nevertheless, the connection between the two models is much closer than we might expect. Purely mathematical theoretical models must still contain useful clinical parameters, while purely clinical models aimed at improving clinical parameters still need mathematical and neurodynamic theory support. The relation between them also raises the question of how to link mathematical theoretical models more closely to clinical predictive models. It can be seen as a "from bench to bedside" problem, i.e., how to bridge the gap between theoretical research and clinical applications so that scientific results can better help human beings.

Moreover, since DBS was originally applied in PD, studying DBS is usually closely linked to studying PD. However, DBS researchers can understand the physiological model of BG in addition to focusing on the pathological BG structure. Translating existing physiological models of BG into DBS models seems to be very promising for research. Due to ethical constraints, researchers have difficulty obtaining DBS data from healthy populations when physiological models of BG may be helpful for the research direction. The assumptions of previous models are constantly disproved because of the continuous development of physiological theories. For example, scholars have discovered the importance of GPe in BG \citep{gittis2014new}. They divided the GPe into two major subpopulations of neurons: prototypic and arkypallidal cells. Recent research results used optogenetic stimulation to study the structure of the GPe \citep{ketzef2021differential}.Rubchinsky found that the chaotic signals can also disrupt the periodic synchronization of neuron \citep{rubchinsky2012intermittent}. McCarthy et al. focused on the rich kinetic phenomena in the striatum\citep{mccarthy2011striatal}. In their latest model DBS can effectively recover the dynamics of striatal networks\citep{adam2022deep}. These novel research results can be well used in computational models. 

Furthermore, because of the good results of DBS for PD, physicians and scholars are now applying DBS to more than 30 other diseases  \citep{Hariz2013}, especially in the group of patients who are poorly controlled despite trying multiple drugs and non-invasive treatments, such as epilepsy, urinary incontinence and depression. In addition, there is increasing academic interest in the mechanisms and optimization of non-invasive stimulation therapies, such as transcranial direct current stimulation, transcutaneous electrical nerve stimulation and transcranial magnetic stimulation. While the benefits of stimulation are clear and robust for some diseases (e.g., PD), there have been some failures in clinical trials for other applications. Furthermore, since the relatively short history of developing DBS treatments for other diseases prevents empiricism from playing a large role, computational models of DBS based on different diseases can be effective in helping to optimize these therapies.

DBS computational models will evolve, perhaps resulting in more concise mathematical formulas or more influential conditions to better match physiological realities. However, in any case, these models can help us uncover more of the brain's secrets. In the future, computational models could act as virtual patients. Doctors and researchers can test new treatment options on the platform and get good results before implementing them on patients.

\section*{Ethics approval}

Written informed consent for publication of this paper was obtained from the South China University of Technology, Ruijin hospital and all authors.

\section*{Disclosure statement}

The authors declare there is no conflicts of interest.

\section*{Funding}

This study was supported by Medical and Engineering Cross Research Fund from Shanghai Jiao Tong University (KH, YG2019QNA31), Shanghai Municipal Health Commission Clinical Study Special Fund (KH, 20194Y0067), Ruijin Youth NSFC Cultivation Fund (KH, 2021$\&$2019), Ruijin Hospital Guangci Excellence Youth Training Program (KH, GCQN-2019-B10), Shanghai Pujiang Program (KH, 19PJ1407500), National Natural Science Foundation of China (Grant Nos.: 11572127 and 11872183).

\section*{Notes on contributor(s)}

First Author and Second Author contribute equally to this work.

\bibliographystyle{tfcse}
\bibliography{DBS}

\end{document}